# Unusual behavior in the first excited state lifetime of catechol

*Martin Weiler †, Mitsuhiko Miyazaki †, Géraldine Féraud §, Shun-ichi Ishiuchi †, Claude Dedonder §, Christophe Jouvet\* § and Masaaki Fujii\* †*

† Chemical Spectroscopy Division, Chemical Resources Laboratory, Tokyo Institute of Technology, 4259-R1-15, Nagatsuta-cho, Midori-ku, Yokohama 226-8503, Japan

§ PIIM−UMR CNRS 7345, Aix Marseille Université, Avenue Escadrille Normandie-Niémen, 13397 Marseille Cedex 20, France

AUTHOR INFORMATION

[*] Corresponding Authors

[*] christophe.jouvet@univ-amu.fr**,** mfujii@res.titech.ac.jp

ABSTRACT: We are presenting vibrationally selective pump-probe measurements of the first electronic excited-state ($\pi\pi^*$) lifetime of jet-cooled neutral catechol (1,2-dihydroxybenzene). The lifetime of the 0-0 transition is very short (7 ps) as rationalized by the small $\pi\pi^*/\pi\sigma^*$ gap calculated. However the lifetimes implying higher out-of-plane vibrational levels are longer (~11 ps). This emphasizes the role of the out-of-plane vibration in the $\pi\pi^*/\pi\sigma^*$ coupling not only in its nature but also in the number of quanta.

**TOC GRAPHICS**

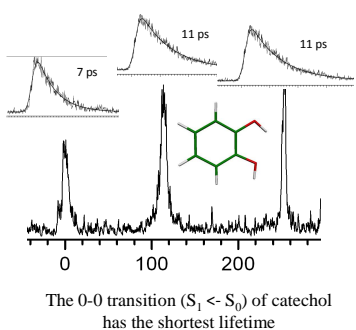

The 0-0 transition ($S_1 \leftarrow S_0$) of catechol has the shortest lifetime

**KEYWORDS:** vibrationally selective pump probe, Hydrogen transfer, non adiabatic coupling, fast decay, aromatic molecules

The excited-state dynamics of aromatic molecules containing an heteroatom (phenol being the prototype molecule) has received considerable attention since it was stated in 2002 that the excited-state dynamics is linked to an Hydrogen atom loss mediated by the tunneling of the H atom from the bound $\pi\pi^*$ to the $\pi\sigma^*$ state dissociative along the OH coordinate.[1] This statement, which was already based on clear experimental[2–4] and theoretical work,[5] has been reinforced by many experimental evidences and calculations.[6–13] In particular, a clear correlation between the $\pi\pi^*$ and $\pi\sigma^*$ energy gap and the $\pi\pi^*$ excited-state lifetime has been measured recently:[14] the smaller the gap, the shorter the lifetime. It should be mentioned that in $C_s$ symmetry (to which ground state phenol and catechol belong) the $\pi\pi^*$ (A') and $\pi\sigma^*$ (A") are not of the same symmetry, the former being A' and the latter A", so that they are not electronically coupled, the coupling occurring through out-of-plane vibrations.[7,13]

The catechol (1,2-dihydroxybenzene) case is quite interesting since it contains two adjacent hydroxyl groups linked by a hydrogen bond. Some unexpected and complicated behavior has been recorded in the product energy distribution.[8] In particular, the energy distribution in the product obtained in detecting the kinetic energy of the H atom is remarkably dependent on the initial $S_1$ out-of-plane vibration.[15]

The excited-state vibrational spectroscopy of the molecule is quite complex and the interpretation has been subject of controversy.[16,17] The agreement is that the low vibrational modes observed are due to out-of-plane OH vibrations and that the excited state has lost the $C_s$ symmetry of the ground state.

A 12 ps excited-state lifetime has been measured with a vibrational excess energy in $S_1$ of $\approx 1800$ cm$^{-1}$.[9] This short lifetime as compared to that of phenol was assigned to the loss of planarity and to the small calculated ππ*/πσ* energy gap (0.29 eV). It was also measured that the $S_1$ lifetimes of resorcinol (1,3-dihydroxybenzene) and of hydroquinone (1,4-dihydroxybenzene) are in nanosecond range (> 1 ns and 0.43 ns respectively)[9] similarly to phenol.[14] These isomers have a larger ππ*/πσ* energy gap (0.51 eV for resorcinol and 0.55 eV for hydroquinone) than catechol.

Very recently, ultrafast time-resolved velocity map ion imaging and time-resolved ion-yield experiments allowed a very detailed analysis of the electronic state relaxation dynamics of photoexcited catechol.[18] In this nice experiment, the excited $S_1$ lifetimes have been recorded on a very broad energy scale from $S_1$ (v = 0) to 5000 cm$^{-1}$ excess energy. Two components have been observed in the decay: one in the 500-900 fs range and one in the 8-10 ps range. The first decay has been ascribed to the IVR in $S_1$ and the second one to the $S_1(\pi\pi^*)/S_2(\pi\sigma^*)$ H atom tunneling. The observation of the fast component on the v = 0 level of $S_1$ is somewhat surprising but it is clear that owing to the laser bandwidth (around 500 cm$^{-1}$), $S_1$ (v = 0) cannot be excited selectively.

The catechol is a chromophore involved in many neurotransmitters and the conformational landscapes of these catecholamines have been recently unraveled in gas phase (supersonic jet) through UV/UV and UV/IR hole burning spectroscopy.[19,20] It has been found that the number of isomers for neurotransmitters containing catechol is much less than for the other chromophores (phenol, resorcinol).[19,20] The possibility of a very short lifetime of some isomers preventing their detection through REMPI method has to be investigated and the measurements of the catechol lifetime is the first step.

We present picosecond pump-probe experiment performed on catechol to see if the number of vibrational quanta observed in the internal distribution of the H-loss product[8] is also observed in the lifetime of the initially excited-state by measuring the excited lifetime of selectively excited vibrational levels.

The origin band at 35 646 cm$^{-1}$ (280.54 nm) is in agreement with what was measured before (35 649 cm$^{-1}$).[17] A low frequency (113 cm$^{-1}$) vibrational progression is observed and was previously characterized.[16,17] The lifetimes have been recorded for selected vibrational levels in this paper. The transient signals for the origin and v=1 of the torsional mode $\tau_1$[8,17] at 113 cm$^{-1}$ are presented in Figure 1 and the fitted lifetime values (from the convolution of a monoexponential decay with 3 ps Gaussian laser width) are gathered in Table 1. Except for the 0-0 transition, all the lifetimes are the same (11 ps) within the experimental errors. They are similar to the ones observed at high vibrational excess energy ($\approx$1800 cm$^{-1}$)[9] and by Chatterley *et al.* on a very broad energy scale.[18] The $S_1$ (v = 0) transition has a significantly shorter lifetime (7 ps) than the excited levels which is not so common. No evidence of a shorter lifetime component could be found, but the laser temporal resolution (3 ps) may not allow such detection.

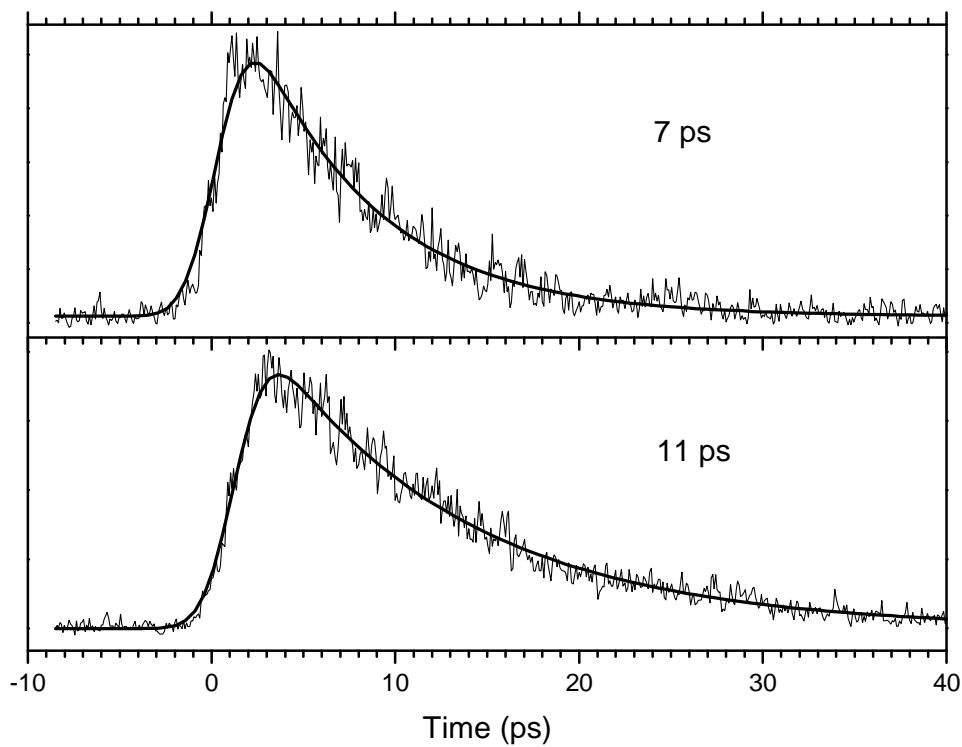

**Figure 1.** Pump-probe delay of catechol when exciting $S_1$ origin (higher panel, thin line) and the vibronic band 113 cm$^{-1}$ above origin (lower panel, thin line). Their fit using a monoexponential decay convoluted by a Gaussian laser width of 3 ps is represented by thick lines.

| Pump energy (cm$^{-1}$) | Pump energy above origin (cm$^{-1}$) | Assignment (from Gerhards et al.[17] and King et al.[8]) | Lifetime (ps) |
|---|---|---|---|
| 35 646 | 0 | $0^0_0$ | 7±0.5 |
| 35 759 | 113 | $\tau^1_{1\,0}$ | 11±1 |
| 35 896 | 250 | $\tau^2_{1\,0}$ | 10±2 |
| 36 378 | 732 | $10b^1_0 16a^1_0$ | 10±2 |

**Table 1**. Lifetimes of the origin and of different vibrational levels in $S_1$ of jet-cooled catechol.

In agreement with previous calculations,[8,9,18] the $\pi\pi^*/\pi\sigma^*$ energy gap at the ground state geometry is very small (0.07 eV) as compared to other phenolic systems.[14] The calculations agree quite well with the experimental data (4.42 eV) for the adiabatic $S_1$ <- $S_0$ transition, calculated at 4.34 eV (including the zero-point energy correction). The optimized $S_1$ excited-state is not planar as already stated in previous papers,[9,16,17] and it is not purely of $\pi\pi^*$ character but has a mixed $\pi\pi^*-\pi\sigma^*$ character. The out of plane deformation involves mainly the both OH (the oxygen are out of plane) but also the H nearby the OH. The carbon skeleton is also out of plane deformed. (see supplementary material the geometries and the vibrations can be viualised using molden or molekel software). The deformation is quite important so that the Cs symmetry rules are not proper to describe the system and that $\sigma$ and $\pi$ nomenclature have to taken as reminiscent of the ground state symmetry but do not apply for the excited state. The same is true for the second excited state, which has a mixed $\pi\sigma^*-\pi\pi^*$ character.

The vibrational modes have been calculated, and the active vibrational modes cannot be assigned only to the OH out-of-plane bending coordinate since the carbon atoms of the aromatic part are strongly involved in the dynamics. The lowest active frequency mode can be seen in Figure 2. (see supplementary material)

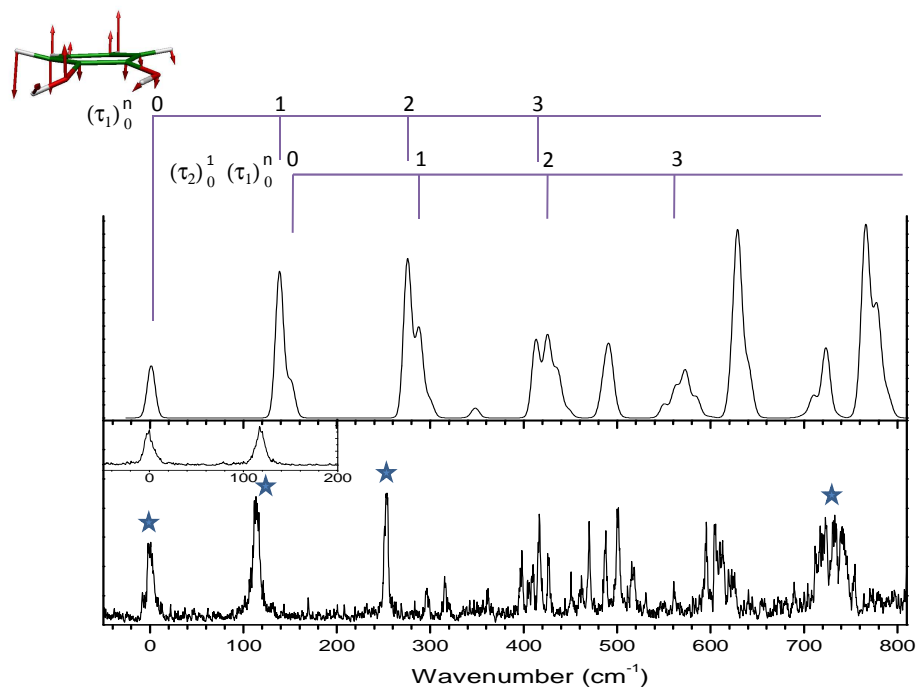

**Figure 2.** Calculated spectrum (upper trace) and comparison with the ns one color REMPI spectrum (lower trace). The vibrations with a star are the ones pumped in the pump-probe measurement. The vibrational assignment based on the two lowest out-of-plane vibrational modes $\tau_1$ (calculated at 137 cm$^{-1}$) and $\tau_2$ (calculated at 150 cm$^{-1}$) is given above the upper trace. In the insert, the spectrum recorded with the picosecond laser shows that the spectral resolution is good enough to excite selectively v=0 and v=1 of the $\tau_1$ mode (observed at 113 cm$^{-1}$).

In the low energy part of the spectrum, the calculated lowest frequency $\tau_1$ (137 cm$^{-1}$) which is responsible for the progression is in reasonable agreement with the observed one (113 cm$^{-1}$). The simulated Franck-Condon envelope is not badly reproduced. As usual the CC2 method tends to increase too much the out-of-plane deformation underestimating the Franck-Condon factor of the 0-0 transition. The barrier calculated for the "inversion coordinate" *i.e.,* when going from the optimized non planar S$_1$ structure to the planar one, can be obtained by optimizing the ππ* state in $C_s$ symmetry. We found a barrier of 240 cm$^{-1}$ which is probably overestimated since the calculated frequency is 15 % higher than the experimental one.

Due to the small ππ*/πσ* energy gap and to the absence of symmetry in the excited-state, one expects a very fast H loss through the ππ*/πσ* tunneling[1] *i.e.,* a short lifetime as observed here and in other experiments.[9,18]

The real unexpected result of the ps pump-probe experiment is coming from the fact that the 0-0 transition has a shorter lifetime than higher vibrational levels. This is not common and has not been observed in the phenol. We can tentatively explain this using a simplified scheme (Figure 3) similar to that of Gerhards *et al.*[17] and King *et al.*[8]

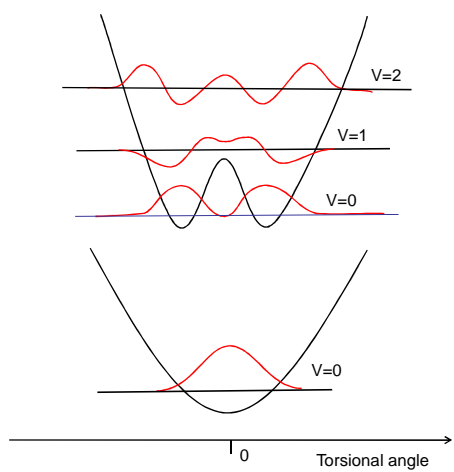

**Figure 3.** artistic view of the wavefunction for the out-of-plane torsional coordinate. For v=0 the probability of presence is non-zero only for the bent geometry, which is not the case for v≥1, so that the ππ*/πσ* coupling is larger for v=0 than for v≥1, possibly explaining the shorter lifetime for the v=0 level adapted from reference 17.

In this picture, the vibrational wavefunctions are presented in a potential along an out-of-plane torsion coordinate, assuming that the barrier is lying between the v=0 and v=1 excited vibronic levels, as calculated. The $\pi\pi^*/\pi\sigma^*$ coupling strongly depends on the out-of-plane deformation, the coupling being essentially zero at the planar geometry.[13] The v=0 level is below the barrier to planarity and thus the vibrational wavefunction is essentially 0 at the planar geometry. Because geometry at v=0 level is always bent, the coupling is quite large and the lifetime should be short. For other vibrational levels the probability of presence in the planar geometry is not zero and the average of the coupling along the vibration is smaller ("the system spends more time in the planar geometry") thus the lifetime should be longer. It would be interesting to see if theoretical calculations can reproduce this experimental result.

The excited-state lifetimes of different vibronic levels in the first excited state of jet-cooled catechol have been measured with ps pump-probe experiment. The lifetimes are very short (11 ps) and even shorter (7 ps) on the 0-0 transition. This can be understood in terms of variation of vibronic coupling between the bound $\pi\pi^*$ and dissociative $\pi\sigma^*$ states due to the absence of $C_s$ symmetry in the excited state. The question of the evolution of this lifetime in substituted catechols and their conformers remains open.

EXPERIMENTAL TECHNIQUES AND CALCULATIONS

Cold catechol is obtained in a molecular beam and investigated using nanosecond and picosecond laser systems, featuring pulse lengths of 5 ns and 3 ps, respectively. Details of the experimental setup, molecule generation, spectroscopic methods, and employed laser systems have been given elsewhere.[21] Briefly, catechol in a supersonic expansion is seeded in 3 bar of Helium and ionized in the extraction region of a time-of-flight mass spectrometer via resonant two-color two-photon ionization (1 + 1' REMPI). The first UV photon excites the molecules into the $S_1$ electronic state, whereas a second UV photon is used for soft ionization with low excess energy. REMPI spectra are obtained by scanning ns UV laser through the vibronic resonances of the $S_1$ state. Lifetime measurement spectra are obtained by fixing the UV wavelength of the ps laser to a particular $S_1$ resonance and scanning the delay of the probe ps laser set at 310 nm, by steps of 0.08 ps.

Ab initio calculations have been performed with the TURBOMOLE program package,[22] making use of the resolution-of-the-identity (RI) approximation for the evaluation of the electron-repulsion integrals.[23] The equilibrium geometry of ground electronic state ($S_0$) has been determined at the RI-MP2[24] level. Excitation energies and equilibrium geometry of the lowest excited singlet state ($S_1$) have been determined at the RI-CC2 level.[25,26] Calculations were performed with the correlation-consistent polarized valence triple-zeta aug-cc-pVTZ basis.[27] The vibrations in the ground and excited-states have been calculated and the vibronic spectra simulated using PGOPHER software[28] for the Frank-Condon analysis.

AUTHOR INFORMATION

**Corresponding Authors**

* Email: christophe.jouvet@univ-amu.fr
* Email: mfujii@res.titech.ac.jp

**Notes**

The authors declare no competing financial interests.

ACKNOWLEDGMENT

This study was supported in part by a Grant-in-Aid for Scientific Research KAKENHI in the priority area (477) "Molecular Science for Supra Functional Systems" from the Ministry of Education, Culture, Sports, Science and Technology (MEXT) Japan and the France-Japan Collaboration Program (SAKURA) France, and the Japan Society for Promotion of Science. MW thanks for Japan Society for Promotion of Science fellowship. We acknowledge the use of the computing facility cluster GMPCS of the LUMAT federation (FR LUMAT 2764)

REFERENCES

(1) Sobolewski, A. L.; Domcke, W.; Dedonder-Lardeux, C.; Jouvet, C. Excited-state hydrogen detachment and hydrogen transfer driven by repulsive πσ* states: A new paradigm for nonradiative decay in aromatic biomolecules. *Phys. Chem. Chem. Phys.* **2002**, *4*, 1093–1100.

(2) Grégoire, G.; Dedonder-Lardeux, C.; Jouvet, C.; Martrenchard, S.; Solgadi, D. Has the Excited State Proton Transfer Ever Been Observed in Phenol–$(NH_3)_n$ Molecular Clusters? *J. Phys. Chem. A* **2001**, *105*, 5971–5976.

(3) Grégoire, G.; Dedonder-Lardeux, C.; Jouvet, C.; Martrenchard, S.; Peremans, A.; Solgadi, D. Picosecond hydrogen transfer in the phenol-$(NH_3)_{n=1-3}$ excited state. *J. Phys. Chem. A* **2000**, *104*, 9087–9090.

(4) Pino, G. A.; Dedonder-Lardeux, C.; Grégoire, G.; Jouvet, C.; Martrenchard, S.; Solgadi, D. Intracluster hydrogen transfer followed by dissociation in the phenol–$(NH_3)_3$ excited state: $PhOH(S_1)$–$(NH_3)_3 \rightarrow PhO^{\bullet}+(NH_4)(NH_3)_2$. *J. Chem. Phys.* **1999**, *111*, 10747.

(5) Sobolewski, A. L.; Domcke, W. Photoinduced Electron and Proton Transfer in Phenol and Its Clusters with Water and Ammonia. *J. Phys. Chem. A* **2001**, *105*, 9275–9283.

(6) Ashfold, M. N. R.; King, G. A.; Murdock, D.; Nix, M. G. D.; Oliver, T. A. A.; Sage, A. G. πσ* excited states in molecular photochemistry. *Phys. Chem. Chem. Phys.* **2010**, *12*, 1218–1238.

(7) Dixon, R. N.; Oliver, T. A. A.; Ashfold, M. N. R. Tunnelling under a conical intersection: Application to the product vibrational state distributions in the UV photodissociation of phenols. *J. Chem. Phys.* **2011**, *134*, 194303.

(8) King, G. A.; Oliver, T. A. A.; Dixon, R. N.; Ashfold, M. N. R. Vibrational energy redistribution in catechol during ultraviolet photolysis. *Phys.Chem.Chem.Phys.* **2012**, *14*, 3338–45.

(9) Livingstone, R. A.; Thompson, J. O. F.; Iljina, M.; Donaldson, R. J.; Sussman, B. J.; Paterson, M. J.; Townsend, D. Time-resolved photoelectron imaging of excited state relaxation dynamics in phenol , catechol , resorcinol , and hydroquinone. *J. Chem. Phys.* **2012**, *137*, 184304.

(10) Roberts, G. M.; Chatterley, A. S.; Young, J. D.; Stavros, V. G. Direct Observation of Hydrogen Tunneling Dynamics in Photoexcited Phenol. *J. Phys. Chem. Lett.* **2012**, *3*, 348–352.

(11) Ishiuchi, S.; Daigoku, K.; Saeki, M.; Sakai, M.; Hashimoto, K.; Fujii, M. Hydrogen transfer in photoexcited phenol/ammonia clusters by UV–IR–UV ion dip spectroscopy and ab initio molecular orbital calculations. I. Electronic transitions. *J. Chem. Phys.* **2002**, *117*, 7077.

(12) Ishiuchi, S.-I.; Daigoku, K.; Hashimoto, K.; Fujii, M. Four-color hole burning spectra of phenol/ammonia 1:3 and 1:4 clusters. *J. Chem. Phys.* **2004**, *120*, 3215–20.

(13) Vieuxmaire, O. P. J.; Lan, Z.; Sobolewski, A. L.; Domcke, W. Ab initio characterization of the conical intersections involved in the photochemistry of phenol. *J. Chem. Phys.* **2008**, *129*, 224307.

(14) Pino, G. A.; Oldani, A. N.; Marceca, E.; Fujii, M.; Ishiuchi, S.; Miyazaki, M.; Broquier, M.; Dedonder, C.; Jouvet, C. Excited state hydrogen transfer dynamics in substituted phenols and their complexes with ammonia: ππ*-πσ* energy gap propensity and ortho-substitution effect. *J. Chem. Phys.* **2010**, *133*, 124313.

(15) Nix, M. G. D.; Devine, A. L.; Cronin, B.; Dixon, R. N.; Ashfold, M. N. R. High resolution photofragment translational spectroscopy studies of the near ultraviolet photolysis of phenol. *J. Chem. Phys.* **2006**, *125*, 133318.

(16) Bürgi, T.; Leutwyler, S. O–H torsional vibrations in the $S_0$ and $S_1$ states of catechol. *J. Chem. Phys.* **1994**, *101*, 8418.

(17) Gerhards, M.; Perl, W.; Schumm, S.; Henrichs, U.; Jacoby, C.; Kleinermanns, K. Structure and vibrations of catechol and catechol $H_2O(D_2O)$ in the $S_0$ and $S_1$ state. *J. Chem. Phys.* **1996**, *104*, 9362.

(18) Chatterley, A. S.; Young, J. D.; Townsend, D.; Żurek, J. M.; Paterson, M. J.; Roberts, G. M.; Stavros, V. G. Manipulating dynamics with chemical structure: probing vibrationally-enhanced tunnelling in photoexcited catechol. *Phys. Chem. Chem. Phys.* **2013**, *15*, 6879–92.

(19) Ishiuchi, S.; Mitsuda, H.; Asakawa, T.; Miyazaki, M.; Fujii, M. Conformational reduction of DOPA in the gas phase studied by laser desorption supersonic jet laser spectroscopy. *Phys. Chem. Chem. Phys.* **2011**, *13*, 7812–20.

(20) Mitsuda, H.; Miyazaki, M.; Nielsen, I. B.; Çarçabal, P.; Dedonder, C.; Jouvet, C.; Ishiuchi, S.; Fujii, M. Evidence for Catechol Ring- Induced Conformational Restriction in Neurotransmitters. *J. Phys. Chem. Lett.* **2010**, *1*, 1130–1133.

(21) Ishiuchi, S.; Sakai, M.; Tsuchida, Y.; Takeda, A.; Kawashima, Y.; Dopfer, O.; Müller-Dethlefs, K.; Fujii, M. IR signature of the photoionization-induced hydrophobic-->hydrophilic site switching in phenol-$Ar_n$ clusters. *J. Chem. Phys.* **2007**, *127*, 114307.

(22) Ahlrichs, R.; Bär, M.; Häser, M.; Horn, H.; Kölmel, C. Electronic structure calculations on workstation computers: The program system turbomole. *Chem. Phys. Lett.* **1989**, *162*, 165–169.

(23) Weigend, F.; Haser, M.; Patzelt, H.; Ahlrichs, R. RI-MP2: optimized auxiliary basis sets and demonstration efficiency. *Chem. Phys. Lett.* **1998**, *294*, 143–152.

(24) Moller, C.; Plesset, M. S. Note on an Approximation Treatment for Many-Electron Systems. *Phys. Rev.* **1934**, *46*, 618–622.

(25) Christiansen, O.; Koch, H.; Jorgensen, P. The 2nd-Order Approximate Coupled-Cluster Singles and Doubles Model CC2. *Chem. Phys. Lett.* **1995**, *243*, 409–418.

(26) Hattig, C. Geometry optimizations with the coupled-cluster model CC2 using the resolution identity approximation. *J. Chem. Phys.* **2003**, *118*, 7751–7761.

(27) Woon, D. E.; Dunning, T. H. Gaussian basis sets for use in correlated molecular calculations . III . The atoms aluminum through argon. *J. Chem. Phys.* **1993**, *98*, 1358–1371.

(28) Western, C. M. PGOPHER, a Program for Simulating Rotational Structure. V 7.0.101.

available at http:// pgopher.chm.bris.ac.uk.